# A CMOS-compatible, scalable and compact magnetoelectric spin-torque microwave detector

Shuhui Liu[1,2,†], Riccardo Tomasello[3,†], Bin Fang[1,2,*], Aitian Chen[4], Like Zhang[5], Zhenhao Liu[1], Rui Hu[1,2], Wenkui Lin[1,2], Mario Carpentieri[3], Baoshun Zhang[1,2], Xixiang Zhang[6], Giovanni Finocchio[7,*], Zhongming Zeng[1,2,*]

[1] Nanofabrication facility, Suzhou Institute of Nano-Tech and Nano-Bionics, Chinese Academy of Sciences, Suzhou, Jiangsu 215123, China

[2] School of Nano Technology and Nano Bionics, University of Science and Technology of China, Hefei, Anhui 230026, People's Republic of China

[3] Department of Electrical and Information Engineering, Politecnico di Bari, I-70125 Bari, Italy

[4] State Key Laboratory of Electronic Thin Film and Integrated Devices, School of Physics, University of Electronic Science and Technology of China, Chengdu, 611731 China

[5] Jiangsu Province Engineering Research Center of Integrated Circuit Reliability Technology and Testing System, Wuxi University, Wuxi, 214105, People's Republic of China

[6] Physical Science and Engineering Division, King Abdullah University of Science and Technology, Thuwal 23955–6900, Saudi Arabia

[7] Department of Mathematical and Computer Sciences, Physical Sciences and Earth Sciences, University of Messina, I-98166, Messina, Italy

[†] *These authors contributed equally:* Shuhui Liu, Riccardo Tomasello

[*]To whom correspondence should be addressed: bfang2013@sinano.ac.cn; gfinocchio@unime.it; zmzeng2012@sinano.ac.cn;

**Abstract**: The development of compact and highly sensitive microwave detectors compatible with complementary metal–oxide–semiconductor (CMOS) processes is an active research area but remains a major challenge in microwave technology. Spin torque diodes (STDs) are emerging nanoscale spintronic devices that have demonstrated a detection sensitivity that surpasses the theoretical thermodynamic limits of Schottky diodes. However, their use for compact microwave detectors is limited by the size of the external antenna or probe. Here, we introduce and demonstrate the concept of a magnetoelectric (ME) spin-torque microwave detector by integrating an ME antenna with a magnetic tunnel junction (MTJ). Our device allows direct electromagnetic-to-DC conversion at sub-microwatt power levels and exhibits a sensitivity of more than 90 kV/W, a noise equivalent power of 3 pW/√Hz, and an area occupancy below 0.4 mm². The high sensitivity is due to the nonlinear coupling between the incoherent magnetization dynamics driven by the DC current applied to the MTJ and both the microwave voltage and strain generated by the incident electromagnetic wave at the ME antenna. We also prove that this detector design is scalable, allowing cointegration of the ME antenna with an array of MTJs. In particular, an ME spin-torque microwave detector with four MTJs exhibits a sensitivity exceeding 400 kV/W. Our work paves the way for a new generation of highly sensitive, compact and scalable microwave detectors that combine ME antennas and spintronic diodes.

**Introduction**

Microwave technologies play critical roles in the fields of communication, physics and medicine, providing essential tools to transmit and receive data over short and long distances (i.e., satellite communication, radar systems, wireless networks, mobile telephones) and for material characterization and medical monitoring[1]. The key components of such technology are microwave detectors. For some highly integrated precision electronic systems for wireless transmission, the electrical characteristics and area occupancy of the microwave detector determine the amount of energy consumed and the overall system size [2,3]. The area occupancy of detectors severely restricts the further development and miniaturization of microwave systems, particularly in applications such as bionic implants and airborne radars[4,5].

Spintronics is an emerging field that promises the development of low-power, compact and high-performance microwave devices such as oscillators[6], amplifiers[7] and detectors[8]. In particular, the development of spin-torque microwave detectors (STMD), such as spin torque diodes (STDs)[9] based on magnetic tunnel junctions (MTJs)[10,11] combined with voltage-controlled magnetocrystalline anisotropy (VCMA)[12] and spin-transfer torque (STT)[13], has shown great potential, offering a nanoscale size, an ultrahigh detection sensitivity (rectified voltage over input microwave power), and a high conversion efficiency at sub-microwatt input powers[14]. These characteristics make STDs promising candidates for the development and deployment of next-generation microwave detectors and electromagnetic energy harvesters[15,16]. On the one hand, conversion efficiencies (delivered DC power over input microwave power) exceeding 5% have been measured in a network of STDs connected in series, for which the enhancement has been attributed to VCMA-driven self-parametric excitation[17,18]. On the other hand, STDs with nonlinear resonance[19] or injection locking[20,21] exhibit good performance in terms of a sensitivity greater than 10 kV/W for nW input powers. Recent studies have demonstrated that the STD sensitivity can reach 1 MV/W when injection locking is coupled with the spin bolometric effect[22]. These previous achievements demonstrated that STDs can surpass the thermodynamic limits of their semiconductor counterparts, such as Schottky diodes[19,20].

However, current STD-based microwave detection systems still require an antenna that relies on electromagnetic wave resonance[23] or a probe to inject a microwave current into the MTJ. Therefore, obtaining an advantage in terms of integration with complementary-metal-oxide-semiconductor (CMOS) processes is challenging despite the nanoscale size and the greater detection sensitivity at low input powers. Thus, further progress is necessary to minimize or eliminate the dependence on large external antennas before STD-based microwave systems can drive the development of ultracompact detectors to reduce the sizes of sensors, radars, etc.

Significant miniaturization of antennas for microwave applications (by 1–2 orders of magnitude) has been obtained with acoustically actuated nanomechanical magnetoelectric (ME) antennas. These devices sense the magnetic and electric fields of electromagnetic waves at their acoustic resonance frequencies, generating an output voltage via the piezoelectric effect[24]. While direct ME rectification has been achieved through approaches such as metamaterial design[25], the use of bulk ME effects[26], and even the development of single organic molecule configurations[27], cointegration of these ME antennas with STDs has not yet been realized. Such integration could represent a key step towards a significant reduction in the size of high-performance and ultralow-power STD-based microwave detectors.

Here, we demonstrate the first proof-of-concept of an ME STMD obtained by cointegrating an ME antenna with MTJ-based STDs. We achieve a rectified voltage directly from the microwave input power delivered by an incident electromagnetic wave to our devices without any additional antenna. Because the ME antenna operation depends on the principle of bulk acoustic wave resonance, an electromagnetic wave signal is detected only when its frequency is within the acoustic resonant frequency band of the ME antenna. In this band, the ME STMD exhibits an ultrahigh detection sensitivity of up to 90 kV/W for sub-microwatt input powers and a noise equivalent power (NEP) of 3 pW/√Hz. The overall 0.4 mm$^2$ area occupancy makes it one of the most compact high-performance microwave detectors developed thus far. Experimental data show that this high sensitivity is related to nonlinear coupling of a large-amplitude dynamic mode excited in the MTJ by a DC electric current and the radiofrequency (RF) excitation resulting from the combined effect of both the microwave voltage and strain of the ME antenna applied to the MTJ[28, 29]. We also demonstrate zero bias field operation. In addition, this ME STMD scheme is highly scalable, allowing for numerous MTJs on top of the ME antenna connected in series to enhance the performance of the ME STMD [17, 18]. We prove a detection sensitivity of up to 446 kV/W for an array of four MTJs connected in series and cointegrated in the same ME antenna without any further increase in the size of the ME STMD. An additional advantage of our device is that all materials used in the ME antenna and the MTJs are CMOS compatible. The detection performance and scalability of our device make it suitable for application as an ultrahigh-performance microwave detector[33] and open a path for the development of a new generation of compact and scalable microwave receivers with high performance in terms of sensitivity and NEP.

**Device description**

The ME STMD is composed of three parts, as sketched in Fig. 1a: (i) an ME antenna realized with an ME heterostructure and a Mo electrode, (ii) an MTJ, and (iii) a Ti/Au coplanar waveguide. Figure 1b shows a transmission electron microscopy (TEM) image of the thin film stack (see Methods "Material deposition and device fabrication") and the energy dispersive spectroscopy (EDS) maps of the ME heterostructure, which highlight the quality of the material interfaces in the ME antenna. The thinner AlN film adjacent to the Si substrate serves as a seed layer, whereas the Mo layer works as the bottom electrode, providing a reliable electrical interface. The ME heterojunction is deposited on top of the Mo layer, and it includes a thicker AlN film acting as a piezoelectric layer with low dielectric loss[34] and an FeGaB/Al$_2$O$_3$ multilayer, which serves as a magnetostrictive layer[35]. An SiO$_2$ insulation layer is deposited on top of the ME antenna film, and then, the MTJ stack, sketched in Fig. 1c, is deposited on top of the SiO$_2$ layer (see Methods "Material deposition and device fabrication" for all details about the film deposition). Figure 1b also shows a TEM image of the whole MTJ stack with a zoomed-in view of the region near the MgO barrier and EDS maps, in which the high quality of the interfaces and the MgO crystal morphology can be clearly observed. All films were deposited on a Si substrate by magnetron sputtering, a CMOS compatible process. Mo and AlN are commonly used in filters, piezoelectric micromachined ultrasonic transducers, and memristors[30, 31, 32]. The FeGaB stack is grown using a process similar to that in CMOS production, while the MTJ uses standard magnetic random-access memory material stacks.

After the films are deposited, the device is fabricated. Figure 1d displays a scanning electron microscopy (SEM) image of the final device that shows how the ME antenna and the MTJ are connected through the Ti/Au coplanar waveguide. The diameter of the fabricated ME antenna is approximately 200 μm, whereas the MTJ is a pillar with an elliptical cross section with $a$ = 440 nm and $b$ = 250 nm as the long and short axes, respectively. The whole device area occupancy is 0.4 mm$^2$. The device size could be further minimized by fabricating the MTJ directly on top of the ME antenna, albeit with increased fabrication complexity. Supplementary Fig. S1 also shows an SEM image of the MTJ. After the nanofabrication is complete, the device is annealed at 300 °C for 1 h while being biased by a 0.8 T magnetic field applied along the -y direction (see the Cartesian coordinate system in Fig. 1e) to align the magnetization of the CoFe (pinned layer) of the synthetic antiferromagnet (SAF) along the -y direction. Once the field is removed, the magnetization of the CoFe/Co$_{40}$Fe$_{40}$B$_{20}$ reference layer (top layer of the SAF) aligns along the +y-axis because of the negative interlayer exchange coupling. This latter magnetization acts as a reference direction onto which the magnetization of the Co$_{40}$Fe$_{40}$B$_{20}$ free layer must be projected to determine the magnetoresistance signal. This magnetic design allows us to set a finite angle between the magnetizations of the free and reference layers under a zero-bias field. This equilibrium angle is given by a trade-off between the shape anisotropy, which tends to align the magnetization along the x-axis, and the dipolar field from the not fully compensated SAF, which is parallel to the y-direction because the CoFe pinned layer gives rise to a dipolar contribution smaller than that of the CoFe/Co$_{40}$Fe$_{40}$B$_{20}$ reference layer.

Figure 1e shows an example of the magnetoresistance as a function of an in-plane magnetic field $H$ applied at $\theta = 15°$ from the positive x-axis. Hereafter, if not otherwise specified, we focus on this field direction, for which the largest detection performance of our ME STMD is achieved. We observe a change in the magnetoresistance of 150 Ω in the field range from -40 mT to 40 mT, which is given mainly by the y-component of the free layer magnetization. We estimate a tunnelling magnetoresistance (TMR), defined as $\mathrm{TMR} = \frac{R_{\mathrm{AP}} - R_{\mathrm{P}}}{R_{\mathrm{P}}} \times 100\%$, of approximately 80% from the measurements of the magnetoresistance loop as a function of the field applied along the y-axis ($\theta = 90°$), as shown in Supplementary Fig. S2. The high-resistance state ($R_{\mathrm{AP}}$) has a resistance of 778 Ω, whereas the low-resistance state ($R_{\mathrm{P}}$) has a resistance of 430 Ω.

Figure 1f shows the $S_{21}$ scattering parameter measured for the ME antenna. The enhanced $S_{21}$ between 400 and 800 MHz (dashed rectangle in Fig. 1f) stems from the acoustic resonance response of the ME antenna. Unlike traditional electric dipole antennas, the ME heterostructure converts incident electromagnetic waves into RF strain and RF current via the magnetostrictive and piezoelectric effects. This process is amplified at the bulk acoustic wave resonance frequencies. Since the direct detection efficiency of electromagnetic waves by the MTJ is very low, the ME antenna is essential for converting incoming radiation into RF signals that are then rectified by the MTJ. We focus on the 400–800 MHz range in the text, where the ME antenna operates with higher efficiency. We also characterized the time response (Supplementary Note 1, Fig. S3) and radiation pattern (Supplementary Note 2 and Fig. S4) of the ME antenna, which exhibits a response speed similar with traditional rod antenna and a unidirectional radiation pattern.

## Results

*Wireless rectification response of the ME STMD*

Figure 2a shows the experimental setup developed to evaluate the rectification response of the ME STMD, which is designed to directly supply a wireless signal to the ME antenna and a DC current to the MTJ and to simultaneously read the rectified voltage $V_{DC}$ across the MTJ terminals. It is composed of a standard gain horn antenna connected to an RF signal generator as the signal source, a DC power supply acting as the current source and a lock-in amplifier to collect the rectified voltage (see also Methods "Wireless detection measurements"). To evaluate the performance of the ME STMD, as a first step, we measure the amount of power delivered to the device ($P_{ME\text{-}antenna}$) as a function of the power generated by the horn antenna ($P_{horn\text{-}antenna}$) with a spectrum analyser (see Supplementary Note 3 and Supplementary Fig. S5). In the remainder of the paper, we consider $P_{ME\text{-}antenna}$ as the power injected into the ME STMD, and we also refer to this as the wireless power.

Figure 2b shows the phase diagram of the rectified voltage $V_{DC}$ as a function of the wireless microwave frequency and the DC current $I_{DC}$ directly applied to the MTJ (from 0.1 to 0.75 mA) for an input $P_{ME\text{-}antenna}$ of 15.1 nW and a bias magnetic field $H$ of 0.7 mT applied along $\theta = 15°$. We observe excitation of three main modes with frequencies $f_1$=540 MHz, $f_2$=570 MHz and $f_3$=592 MHz, with the largest $V_{DC}$ in the range of positive currents of 0.55 mA ~ 0.75 mA (See Supplementary Note 4 for a detailed explanation of the spectral features in the rectification response). Figure 2c displays the rectification curves obtained for $I_{DC}$ = 0.65 mA as a function of the microwave frequency at different $P_{ME\text{-}antenna}$ values, whose qualitative behaviour is almost independent of $P_{ME\text{-}antenna}$ from 0.7 nW to 24.5 nW. This behaviour depends on the combination of the acoustic response of the ME antenna and the rectification properties of the MTJ, as we explain in the next section. Figure 2d shows the $V_{DC}$ at 540 MHz as a function of $P_{ME\text{-}antenna}$. Interestingly, $V_{DC}$ exceeds 2.0 mV at $P_{ME\text{-}antenna}$ = 24.5 nW. Figure 2d also summarizes the detection sensitivity $S$ at 540 MHz as a function of $P_{ME\text{-}antenna}$. $S$ monotonically increases in this power range, approaching 91.9 kV/W for $P_{ME\text{-}antenna}$= 19.3 nW, and then saturates at greater powers once the magnetization precession saturates, as already observed in the previous STD response[20]. All measurements were performed under non-ideal impedance matching conditions.

We also demonstrate good wireless detection performance for a range of fields, including zero-field operation of our ME STMD, with a response qualitative similar with that observed in Fig. 2b and a sensitivity of 40 kV/W at 24.5 nW (see Supplementary Note 5 and Supplementary Figs. S6 and S7). From a technological perspective, an external field is typically provided by additional hard magnets or supplied current lines. Therefore, designing a detector that operates at zero field is highly relevant because the overall complexity and power requirements of the system are significantly reduced. In other words, zero-field operation can minimize energy consumption, enhance scalability and simplify the strategy for future full on-chip cointegration with CMOS technology.

We also characterize the signal-to-noise ratio (SNR) of the ME STMD by measuring an NEP of approximately $3 \times 10^{-12}$ W/√Hz (see Supplementary Note 6), which is close to the value of a conventional CMOS diode[19]. This characteristic enhances the competitiveness of our device.

In summary, the main result of this work is that our ME STMD has (i) an ultrahigh sensitivity > 90 kV/W at its resonance frequency, (ii) a compact design with an area occupancy of less than 0.4 mm$^2$, which is affected mainly by the size of the ME antenna, (iii) a working input power at the nW level[19, 20], (iv) field-free operation and (v) a small NEP.

## Working principle of the ME STMD

In this section, we describe the origin of the rectification characteristics of our ME STMD design by performing additional experiments to identify and disentangle the different contributions.

*Effect of the DC current.* To understand the role of the injected DC current, we measure the microwave emissions as a function of $I_{DC}$ at zero $P_{ME-antenna}$. Examples of the microwave spectra are summarized in Fig. 3a for $I_{DC}$ = 0.35, 0.45, 0.55 and 0.75 mA. We observe excitation of a dynamic state characterized by a mode with a broad linewidth. In other words, the DC current excites incoherent magnetization dynamics (qualitatively confirmed by micromagnetic simulations; see Supplementary Note 7 and Supplementary Fig. S9a) because of the large size of the device, which favours excitation of nonuniform modes. In addition, the DC current range within which these dynamic modes are excited coincides with that within which a large response of the ME STMD is measured (see Fig. 2b), meaning that the DC current significantly contributes to the wireless detection performance. With this in mind, we verify the effect of the application of a microwave signal in this dynamic state by measuring the microwave emissions in the presence of both $I_{DC}$ and $P_{ME-antenna}$ up to 30 nW and observe that there is no injection locking, as in previous devices[20]. Therefore, we conclude that the mechanism driving the large rectification of our device has a different origin, as discussed below.

*Effect of strain.* Considering the previous result and to further understand the behaviour of the ME STMD,

we also fabricate additional devices in which the ME antenna and the MTJ are disconnected (the Ti/Au coplanar waveguide is removed), and an RF current together with a DC current can be directly applied to the MTJ (see Fig. 3b for the measurement system used and the inset of Fig. 3c for an optical image of the device). Figure 3c shows the rectification curves for a DC current of 0.33 mA at a field of −2.3 mT applied along $\theta$ = 30° under three different conditions: (i) direct RF power of 34 nW applied to the MTJ via a probe, as in standard ST-FMR measurements (blue curve, no input to the horn antenna), (ii) electromagnetic wave delivering $P_{ME\text{-}antenna}$ = 12 nW (red curve, no input to the attenuator), and (iii) both sources applied simultaneously (green curve, configuration shown in Fig. 3b). In (i), the MTJ does not exhibit any rectification response because the RF input power is not large enough to generate a measurable rectified voltage. In (ii), the MTJ exhibits an overall response qualitative similar with that in Fig. 2, with rectification peaks at $f_1$ = 529 MHz and $f_2$ = 692 MHz having voltages of $V_{DC1}$ = -73.4 μV and $V_{DC2}$ = -46.7 μV and sensitivities of $S_1$ = 6117 mV/mW and $S_2$ = 3892 mV/mW, respectively, and the negative rectified voltages are due to the different measurement conditions compared with those in Fig. 2. Since there is no effective electrical connection between the MTJ and the ME antenna, the rectified voltage cannot be attributed to direct excitation of the free layer magnetization by the microwave voltage given by the piezoelectric material. The rectification curves measured in the disconnected device highlight the presence of an effective mechanism for excitation of the MTJ, which can be related to the strain-mediated magnetoelectric coupling (SMMC) mechanism[24, 26]. When the frequency of the electromagnetic wave applied to the ME antenna matches the acoustic resonance frequency of the device, a large strain at the same microwave frequency is excited. This design of the ME STMD enables direct application of strain from the ME substrate to the MTJ via the $SiO_2$ layer (see Fig. 1a). The data depicted in Fig. 3b measured under condition (iii) (i.e., a microwave current and an electromagnetic wave simultaneously applied) show a significant enhancement of the rectification response, confirming that the strain plays an active role in the high-sensitivity response of our ME STMD. The SMMC induces modulation of the anisotropy of the device that changes the oscillation axis of the magnetization precession, similar to what has already been observed for nonvolatile ME switching of MTJs[33-36, 37] (see also Supplementary Note 7 and Supplementary Fig. S9d for the micromagnetic description of the strain in the Landau–Lifshitz–Gilbert (LLG) equation).

*Quantification of the rectification response.* To quantitatively analyse the data from the ME STMD, we consider a model that describes the rectified voltage $V_{DC}$ of STDs working in a nonlinear regime. This is given by two components, as described by the following equation:

$$V_{\text{DC}} \approx \frac{1}{2} I_{\text{RF-M}} \Delta R_S \cos\varphi_s + I_{\text{DC}} \Delta R_{\text{DC}}. \tag{2}$$

where $\Delta R_{\text{DC}} = R_{\text{DC}}(P_{\text{ME-antenna}}) - R_{\text{DC}}(P_{\text{ME-antenna}} = 0)$ is the variation in the DC resistance driven by the microwave excitation[20, 38], $I_{\text{RF-M}}$ is the amplitude of the RF current applied to the MTJ, $\Delta R_S$ is the oscillating component of the magnetoresistance at the same frequency as the microwave current, and $\varphi_s$ is the phase shift between the input microwave current and the oscillating magnetoresistance. The first term is related to the linear response of the ME STMD; $I_{\text{RF-M}}$ can be evaluated by the input power delivered to the MTJ from the ME antenna, whereas $\Delta R_S \cos\varphi_s$ can be evaluated from the power of the microwave emissions at that frequency. For example, at $I_{\text{DC}}$ = 0.65 mA and $I_{\text{RF-M}}$ = 6.6 µA ($P_{\text{ME antenna}}$ = 24.5 nW), Fig. 3d gives $\Delta R_{\text{DC}}$ = 3 Ω, yielding a DC voltage of 1.95 mV. The linear component $\frac{1}{2} I_{\text{RF-M}} \Delta R_S \cos\varphi_s$ is estimated as follows[20]: $\Delta R_S$ = 17 Ω from the microwave emissions data under combined DC and RF excitation, and $I_{\text{RF-M}}$ related to the voltage generated by the ME antenna and supplied to the MTJ, is estimated to be < 6.6 µA, giving an amplitude of ~ 56 µV, one order of magnitude smaller than the non-linear term $\Delta R_{\text{DC}} I_{\text{DC}}$. Given that non-zero phase $\varphi_s$ suggested by the precession mode in Fig. 3a, the actual linear component is likely even smaller. Hence, the nonlinear term is the dominate contribution of the rectified voltage.

The second term can be calculated by directly measuring $\Delta R_{\text{DC}}$, which, in this device, is determined by the coupling between the incoherent magnetization dynamics due to the injected DC current (Fig. 3a) and microwave voltage and the SMMC generated by the ME antenna. The latter drives a variation in the DC resistance $R_{\text{DC}}$ by changing the oscillation axis of the magnetization dynamics. Figure 3c shows $R_{\text{DC}}(0)$ with no microwave input ($P_{\text{ME-antenna}}$ = 0 nW) and with $P_{\text{ME-antenna}}$ = 24.5 nW for $f$ = 540 MHz and 592 MHz as a function of $I_{\text{DC}}$. The three curves exhibit qualitatively similar behaviours, but a finite value of $P_{\text{ME-antenna}}$ quantitatively changes the DC resistance. An example of a direct comparison of the experimental $V_{\text{DC}}$ with the $V_{\text{DC}}$ computed from the nonlinear term of Eq. 1 ($I_{\text{DC}} \cdot \Delta R_{\text{DC}}$), directly estimated from the experimental data, as a function of $I_{\text{DC}}$ is shown in Fig. 3d ($P_{\text{ME-antenna}}$ = 24.5 nW, $f$ = 540 MHz). The results are in very good quantitative agreement, confirming that the nonlinear response of the MTJ due to the coupling between the incoherent dynamic state and microwave voltage and the SMMC is the key ingredient driving the large rectification observed in our ME STMD.

*Working principle of the ME STMD.* Now that we have disentangled all the contributions and deeply understood the rectification mechanism, we summarize the complete working principle of our STMD, which is sketched in Fig. 3e. The magnetic and electric field components of an incident electromagnetic wave couple with the magnetostrictive layer (FeGaB/Al$_2$O$_3$) and the piezoelectric layer (AlN) of the ME antenna and drive two excitations of the MTJ. The first excitation corresponds to a direct microwave voltage due to the piezoelectric layer, whereas the second excitation corresponds to the microwave strain, which couples the magnetic and electric parts of the ME antenna. When the MTJ is biased with a DC current large enough to drive excitation of incoherent dynamics, a large rectification diode effect is measured because of the coupling of these dynamics with the two excitation mechanisms driven by the ME antenna.

*Scalable integration properties*

To demonstrate the scalability of our ME STMD, we realize additional samples using the same film stack but with multiple MTJs connected in series on top of the ME substrate. The array of STDs is then electrically connected to the ME antenna via a Ti/Au coplanar waveguide. An example of a device with four MTJs is shown in Fig. 4a, in which the ME antenna is on the right side and the series-connected array of MTJs is on the left side. Figure 4b shows the magnetoresistance as a function of the external in-plane field applied at $\theta = 90°$ for devices with different numbers of MTJs from 2 to 4. The TMR ranges from 80 to 90% depending on the device. For instance, the TMR for the ME STMD with four MTJs is 87.1%.

Figure 4c shows the maximum detection sensitivity as a function of the number of MTJs. For $P_{\text{ME-antenna}}= 24.5$ nW, the sensitivity of the ME STMD with four MTJs in series reaches 446 kV/W, and the results show a linear increase in sensitivity with the number of MTJs. The operating frequencies of the devices with different numbers of MTJs are still in the range of 510-650 MHz, as shown by a comparison of the panels in Fig. 4d. However, to obtain the best rectification performance in each device, the magnetic field and bias current are slightly adjusted across different devices during the testing phase. As shown in Fig. 4d, while the frequency of the main rectification peak remains almost unchanged, the minor rectification peaks exhibit slight variations as the number of devices increases. This behaviour could be attributed to the slight device-to-device variations due to the nanofabrication process.

Figure 4e-g shows the phase diagram of the rectified voltage for the STMD with 2, 3, and 4 MTJs connected in series as a function of $I_{\text{DC}}$ for $P_{\text{ME-antenna}}= 24.5$ nW. The larger rectification peak remains at approximately 529 MHz, indicating that the scalable approach can potentially be suitable for larger-scale fabrication. Furthermore, the DC current required for achieving a greater detection sensitivity progressively decreases with increasing number of series-connected MTJs. This result is similar to that observed in a previous

work on an array of MTJs working for electromagnetic energy harvesting[18] and can be related to both electrical and magnetic coupling among the MTJs[39]. Our results show a scalable approach that allows for a significant enhancement in the wireless detection performance while keeping the area occupancy constant.

**Discussion**

As described above, we prove that our ME STMD design can be used to reduce the size of microwave detectors in many applications for which a small area occupancy is a major requirement while maintaining high detection performance and nW power detection. To prove the significance of our results, we compare the characteristics of our ME STMD with those of state-of-the art microwave detectors in Table 1.

We finally wish to highlight that the working principle of our ME STMD differs from those of previously reported devices, which are based on nonadiabatic stochastic resonance[38], nonlinear resonance[19], vortex expulsion[40], or injection locking[20], and can be described by the same simple mathematical model given by Eq. 1. As a prospective research direction, the MTJ of our ME STMD could be replaced with STDs on the basis of previously observed mechanisms[19, 20, 38, 40].

Table 1 | Comparison with other similar microwave detector devices

| Ref. | Sensitivity (V/W) | Bias current | Need extra Antenna | Dynamic range(dB) | NEP (W/$\sqrt{Hz}$) | Sub-µW detection | Technology |
|---|---|---|---|---|---|---|---|
| [41] | 3800 | No | Yes | ~30 | $1.9\times10^{-12}$ | No | SMS 7630 |
| [42] | $1.58\times10^4$ | No | Yes | / | / | No | SMS7630+voltage doubler+print antenna |
| [43] | $2\times10^4$ | Yes | Yes | ~30 | ~$10^{-12}$ | No | DDB 2503 |
| [19] | $1.2\times10^4$ | Yes | Yes | ~20 | $3.6\times10^{-12}$ | Yes | Single MTJ diode |
| [17] | $2\times10^4$ | Yes | Yes | / | / | Yes | MTJ array + patch antenna |
| This work | $4.4\times10^5$ | Yes | No | > 15 | $6\times10^{-12}$ | Yes | Integrated MTJ array + ME antenna |

**Conclusions**

In summary, we have reported the first proof-of-concept of a compact (area occupancy <0.4 mm$^2$) and scalable ME STMD able to detect input signals at the nW power level with a high detection sensitivity, a small NEP, and the possibility of operating without a magnetic field. In particular, our experimental findings indicate a wireless detection frequency range of 510–650 MHz with detection sensitivities of approximately 90 kV/W and 400 kV/W for the ME STMD with a single MTJ or an array of 4 MTJs, respectively. We have proven that the design of our device cointegrating an ME antenna and an STD-based MTJ allows coupling of the incoherent magnetization dynamics excited in the MTJ with the RF voltage and strain from the ME antenna. Thus, it is the key element driving the high sensitivity. Further strategies will require cointegration with ME antennas having broader detection bands[44] to achieve wide-band microwave collection or a film bulk acoustic resonator (FBAR)-based antenna to achieve a much higher detection sensitivity.

Our work not only opens the path for the use of the SMMC mechanism as a novel method to further improve the detection sensitivity of STDs but also represents a more compact solution than state-of-the art commercial detectors, thus promising a strong impact on the development of next-generation microwave technology, paving the way for hybrid ME-spintronics technology.

## Methods

### Material deposition and device fabrication

The film stack comprises a Si substrate/AlN (50)/Mo (200)/AlN (1000)/[FeGaB (55)/Al$_2$O$_3$ (5)]$_5$/FeGaB (55)/SiO$_2$ (300)/Ta (5)/Ru (10)/Ta (5)/Co$_{40}$Fe$_{40}$B$_{20}$ (1.2)/MgO (0.85)/Co$_{40}$Fe$_{40}$B$_{20}$ (2.0)/Co$_{70}$Fe$_{30}$ (0.5)/Ru (0.85)/Co$_{70}$Fe$_{30}$ (2.5)/IrMn (8)/Ta (5)/Ru (20) (the number in parentheses indicates the thickness in nanometres). To achieve a smooth surface for subsequent deposition of the MTJ stack, a SiO$_2$ insulation layer was grown above the ME antenna film using inductively coupled plasma chemical vapour deposition (ICPCVD) equipment (Oxford Plasma System 133), and then, a chemical mechanical polishing (CMP) process was performed to reduce its surface roughness to 0.3 nm. The MTJ stack, sketched in Fig. 1c, was then deposited on top of the SiO$_2$ layer. The MTJ structure has an SAF (CoFe/Ru/CoFe/Co$_{40}$Fe$_{40}$B$_{20}$) exchange biased with an IrMn layer designed to fix the direction of the magnetization of the CoFe pinned layer of the MTJ device. The free layer is composed of a 1.2 nm Co$_{40}$Fe$_{40}$B$_{20}$ film separated from the reference layer CoFe/Co$_{40}$Fe$_{40}$B$_{20}$ by a 0.9 nm MgO tunnel barrier. The films of the MTJ and ME antenna were deposited by a magnetron sputtering process (Singulus ROTARIS).

The films were patterned using optical and electron-beam lithography (Raith 150) combined with Ar-ion beam milling (Scia mill 200) and inductively coupled plasma etching (NAURA-GSEC200, for AlN etching). The sidewalls of the ME antenna and MTJ were insulated with SiN$_x$ (deposited with an ICPCVD process at 75 °C), and then, the electrode vias were fabricated using reactive ion etching (RIE) with Oxford 80+ equipment. A Ti (20 nm)/Au (120 nm) film was deposited at the end of the fabrication process as the coplanar waveguide (with a magnetron sputtering process).

**Wireless detection measurements**

A probe station was used for rectification measurements. The wireless rectification performance was measured with the spin-torque ferromagnetic resonance (ST-FMR) technique. An RF signal modulated at a low frequency of 117 Hz was applied by a lock-in amplifier (Stanford SR830), and the resulting rectified DC voltage was measured by the lock-in amplifier with a time constant of 100 ms and a low-pass filter slope of 24 dB per octave. A DC current was applied by a Keithley 2450 power meter. $S_{21}$ was measured using an R&S ZVA40 vector network analyser. During the test of the separate device in Fig. 3b, the signal generated by the RF signal generator was divided into two RF signals ($RF_1$ and $RF_2$) by a directional coupler, $RF_1$ was injected into the horn antenna, and $RF_2$ was attenuated by an attenuator and then directly injected into the device. A bias tee was used to isolate the input RF signal from the rectified voltage ($V_{DC}$) and the DC bias applied to the device ($I_{DC}$).

**Acknowledgements**

The work was supported by the National Natural Science Foundation of China (NNSFC) (Nos. 52371206, 12474127, 12204357, and U24A6001). This study was partially supported by the CAS Young Talent Program and the Gusu Leading Talents Program (No. ZXL2023172). A.C. acknowledges support from the National Key Research and Development Program of China (No. 2024YFA1408503) and the Sichuan Province Science and Technology Support Program (No. 2025YFHZ0147). The work of G.F. was supported by the MUR-PNRR project SPINERGY "SPINtronic Electromagnetic eneRGY harvesting with magnetic tunnel junctions for next generation of green IoT nodes", CUP D93C22000900001 by Nest – Network 4 Energy Sustainable Transition, Parternariato Esteso – PE000002. The work of R.T. and M.C. was partially supported by the project PRIN20222N9A73 "SKYrmion-based magnetic tunnel junction to design a temperature SENSor—SkySens", funded by the Italian Ministry of Research, and by the Project PE0000021, "Network 4 Energy Sustainable Transition – NEST", funded by the European Union – NextGenerationEU, under the National Recovery and Resilience Plan (NRRP), Mission 4 Component 2 Investment 1.3 - Call for tender No. 1561 of 11.10.2022 of Ministero dell'Università e della Ricerca (MUR) (CUP C93C22005230007). R.T., M.C., and G.F. are with the Petaspin TEAM and are thankful for the support of the PETASPIN association (www. petaspin.com).

## Data statement

The data that support the findings of this study are available from the corresponding author upon reasonable request.

## Conflict of interest

The authors declare that they have no conflicts of interest.

## Author contributions

B.F., Z.Z. and G.F. designed the experiments. R.H., W.L., A.C. and X.Z. prepared the films. S.L. performed the device fabrication and TEM characterization. S.L. and B.F. performed the electrical characterizations. R.T., M.C. and G.F. designed the micromagnetic solver. R.T. and M.C. carried out the micromagnetic simulations. B.F., S.L., Z.L., L.Z., B.Z. and Z.Z. analysed the data. B.F., S.L., G.F. and R.T. wrote the manuscript with the help of Z.Z. All the authors commented on the final version of the manuscript. The work was performed under the supervision of B.F., Z.Z. and G.F.

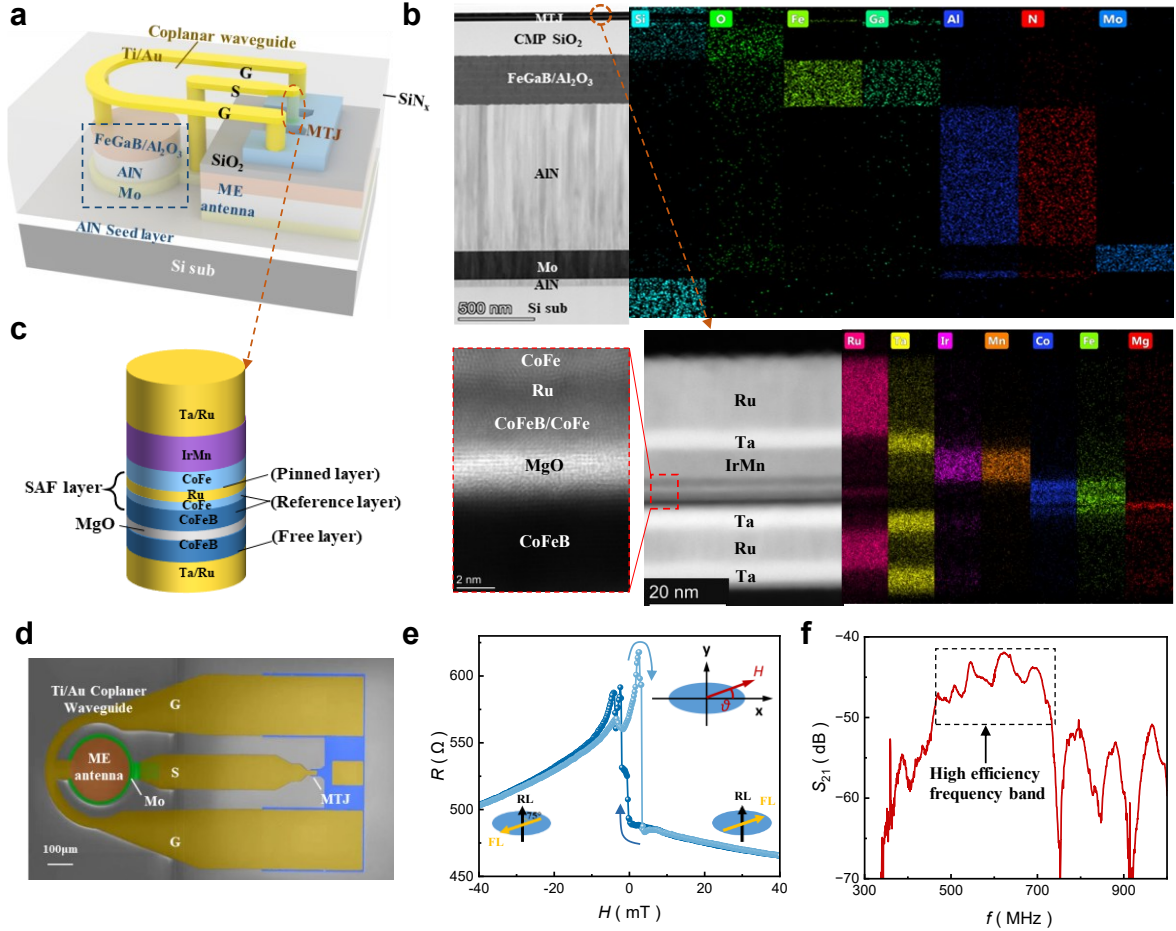

**Figure 1 | Characteristics of the device. a,** Schematics of the ME STMD with the ME antenna, MTJ and Ti/Au coplanar waveguide indicated. The magnetostrictive layer (FeGaB/Al$_2$O$_3$), the piezoelectric layer (AlN) and a bottom electrode (Mo) together form the ME antenna. FeGaB/Al$_2$O$_3$ also acts as the top electrode of the antenna. The MTJ and ME antenna are connected by a Ti/Au electrode realized during device fabrication. **b,** TEM image and EDS maps of the device stack and the MTJ multilayer. The thinner AlN film at the bottom is used as a seed layer. SiO$_2$ undergoes a chemical mechanical polishing process to ensure the quality of the MTJ film. **c,** Sketch of the device structure of the MTJ. **d,** SEM image of the fabricated ME STMD. The yellow region is the Ti/Au electrode. The orange region is the FeGaB/Al$_2$O$_3$ and AlN layers. The green region is the bottom electrode of the ME antenna (Mo). The blue region is the MTJ bottom electrode. **e,** Magnetoresistance curve of the MTJ device under an in-plane magnetic field $H$ applied along $\theta = 15°$ with respect to the positive x-axis. The black arrow in the ellipse indicates the direction of the CoFe/Co$_{40}$Fe$_{40}$B$_{20}$ reference layer (RL) magnetization, whereas the yellow arrow indicates the direction of the CoFeB free layer (FL) magnetization. The Cartesian coordinate system is also shown. **f,** S$_{21}$ scattering parameter for the ME antenna measured up to 1 GHz. The band with a high detection efficiency is within the frequency range between 400 and 800 MHz.

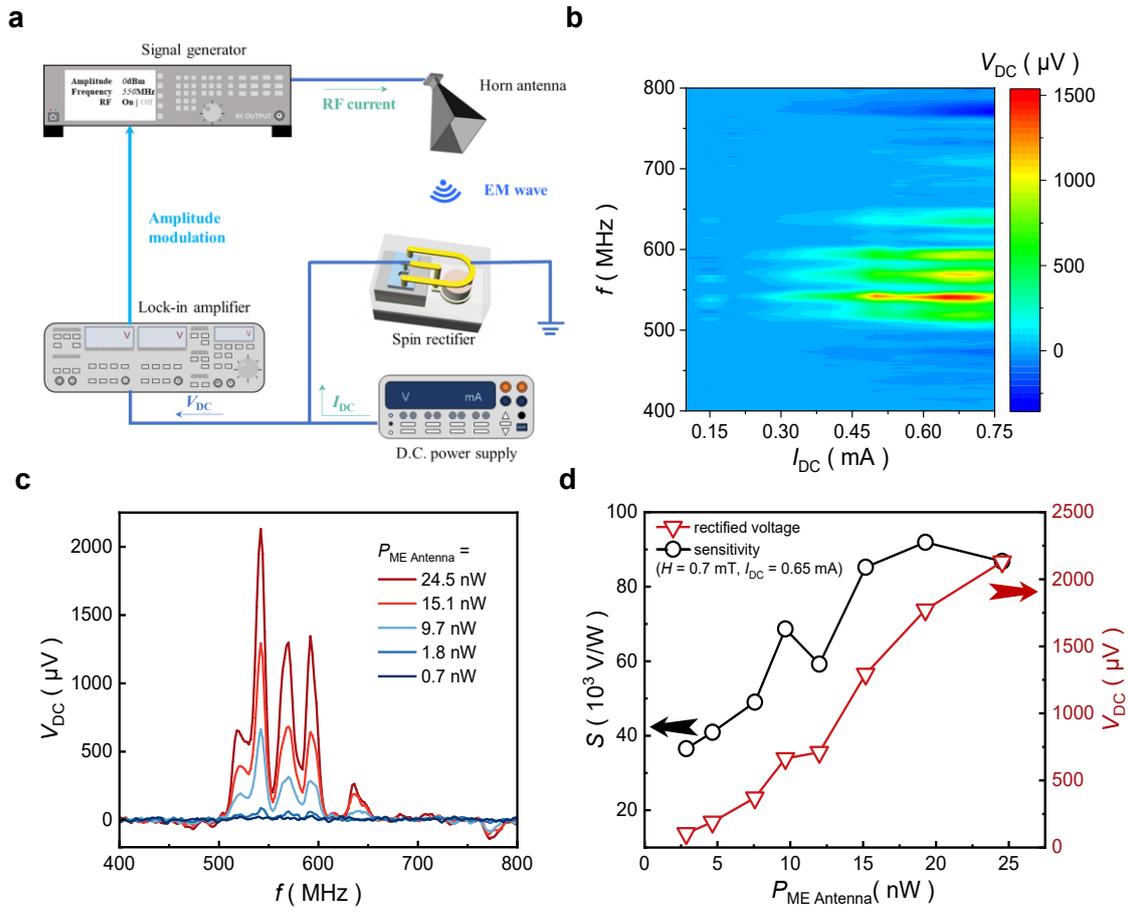

**Figure 2 | Wireless detection performance of the device. a,** Schematic of the circuit used for wireless rectification measurements. A microwave signal is generated by a high-frequency signal generator and sent to the device through a horn antenna. The rectified voltage $V_{DC}$ is collected by a lock-in amplifier. **b,** Wireless rectification performance at different DC currents. The main working frequency range of the device is between 510 and 650 MHz, and the maximum rectified voltage is obtained at $I_{DC}$ = 0.65 mA. **c,** Wireless rectification performance for different $P_{\text{ME-antenna}}$ values. **d,** Detection sensitivity (black line) and rectified voltage (red line) as a function of $P_{\text{ME-antenna}}$. The data are measured at $H$ = 0.7 mT and $I_{DC}$ = 0.65 mA.

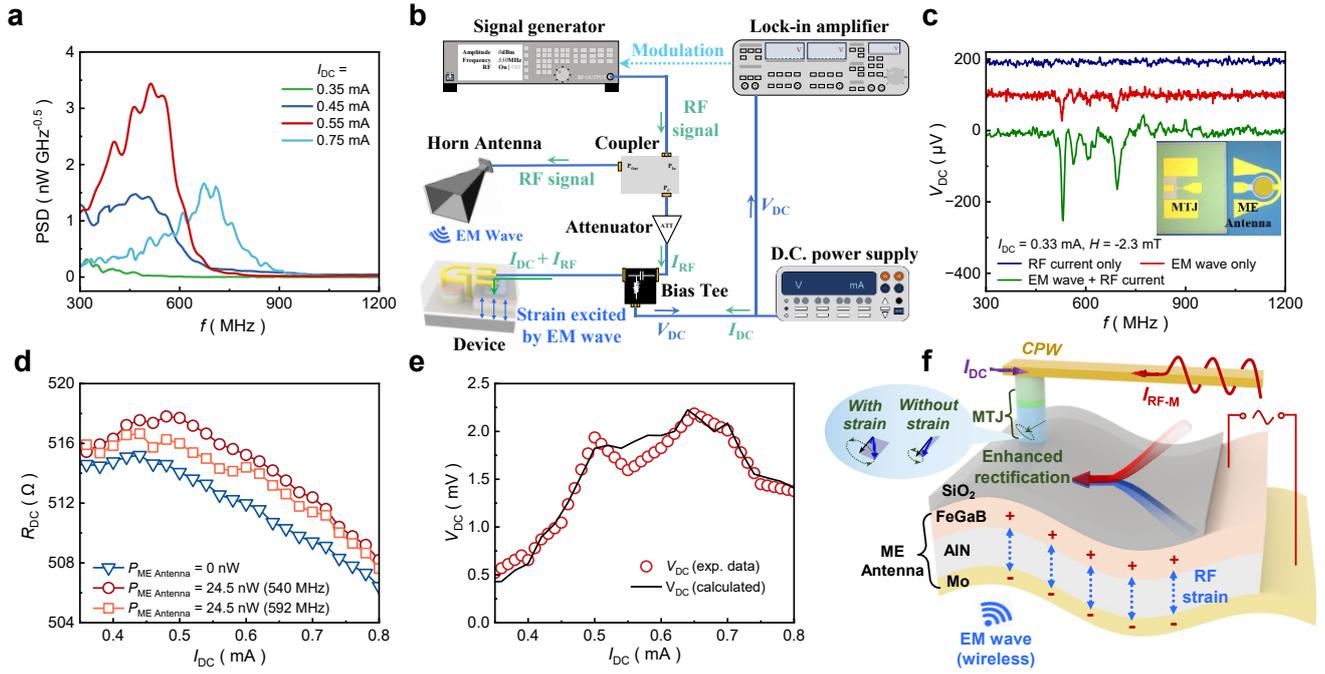

**Figure 3 | Operation Mechanism of the ME STMD. a,** Power spectral density (PSD) of microwave emissions at different $I_{DC}$, showing excitation of nonuniform precession modes with a large bandwidth. **b,** Test circuit for the disconnected device. The ME antenna does not supply RF electrical input to the MTJ. A coupler splits the signal: most goes to the horn antenna (converted to EM wave), a small fraction is attenuated and injected into the MTJ. **c,** Rectification response when the ME antenna is disconnected from the MTJ ($I_{DC}$ = 0.33 mA and $H$ = -2.3 mT) under three conditions: direct RF power (34 nW) to the MTJ (blue), EM delivering ($P_{ME\text{-}antenna}$ = 12 nW) (red), and both sources applied (green). The combined case shows enhanced rectification due to electrical and strain-mediated magnetoelastic coupling (SMMC). The inset shows the optical image of the disconnected device. **d,** DC resistance $R_{DC}$ vs. $I_{DC}$ with and without wireless RF signal ($P_{ME\text{-}antenna}$ = 24.5 nW) at $f$ = 540 MHz (red circle dot) and 592 MHz (orange square dots). **e,** Comparison of experimental rectified voltage (red dots) and the calculation from the nonlinear term of Eq. 2 (solid line). **f,** Working principal schematic: under an EM wave, the AlN layer generates an RF current delivered to the MTJ, while the ME heterostructure produces a microwave strain that modulates the free-layer anisotropy via SMMC. Rectification is enhanced by the combined action of RF voltage, RF SMMC, and incoherent dynamics excited by the DC current.

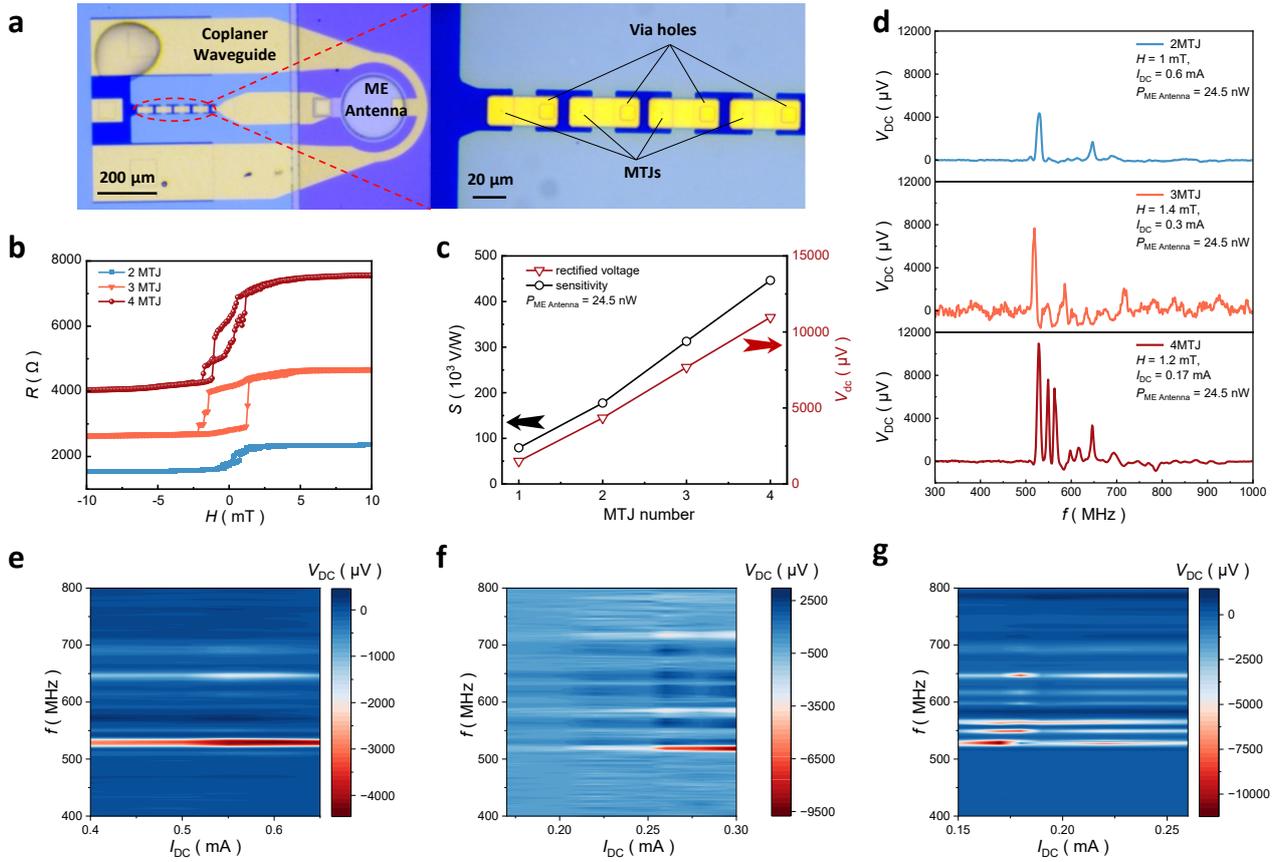

**Figure 4 | Detection performance of devices with multiple MTJs in series. a,** Optical image of the ME STMD with four MTJs. **b,** Magnetoresistance characteristics of the ME STMD with different numbers of MTJs in series as a function of an in-plane field applied at $\theta = 90°$. The intermediate step resistance states are due to the different coercive fields of the MTJs determined by the small device-to-device variations in the physical and geometrical parameters. The sweep step of the magnetic field is 0.1 mT. **c,** Maximum sensitivity and rectified voltage as a function of the number of MTJs. **d,** Rectified voltages of devices with different numbers of MTJs in series, as indicated in the figure legend, for enhanced wireless detection. **e-g,** Rectified voltage as a function of $I_{DC}$ and the frequency of the wireless power for $P_{\text{ME-antenna}} = 24.5$ nW: **e,** 2 MTJs, **f,** 3 MTJs and **g,** 4 MTJs in series.

# Supplementary Information for

# A CMOS-compatible, scalable and compact magnetoelectric spin-torque microwave detector

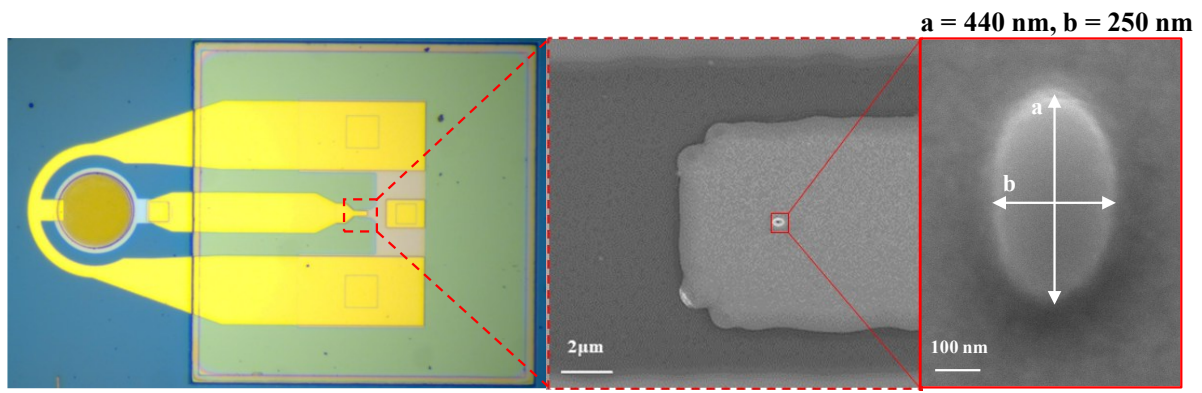

**Supplementary Figure S1.** A detailed SEM image of the device with a zoom in the region where the MTJ is realized. The SEM of the MTJ is also shown. The device is patterned into an elliptical cross section with the major and minor axis having a nominal value of 440 nm and 250 nm, respectively.

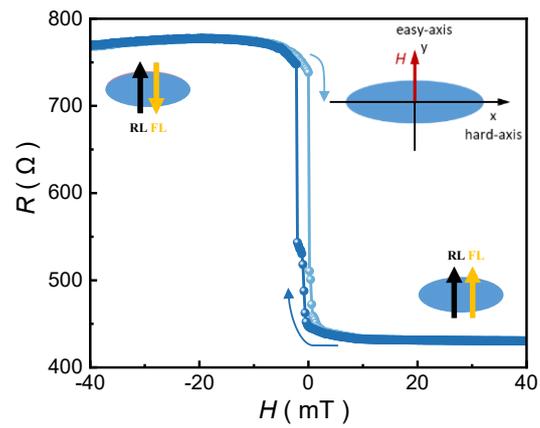

**Supplementary Figure S2.** The magnetoresistance field scan obtained for an in-plane field $H$ applied at $\theta = 90°$ used to estimate the tunneling magnetoresistance ratio. The black arrow in the ellipse indicates the direction of the reference layer (RL) magnetization while the yellow one indicates the direction of the free layer magnetization (FL).

**Supplementary Note 1 - Time response of the ME antenna.**

The time response of the ME antenna is evaluated using the measurement scheme displayed in Supplementary Fig. S3a. A pulse sin wave ($T_{pulse}$ = 200 ns) with a specific frequency (540 MHz) is sent to the horn antenna by an arbitrary waveform generator (Tektronix, AWG7122C). The RF signal provided by the AWG was divided into two RF signals ($RF_1$ and $RF_2$) by a directional coupler (16dB). $RF_1$ and $RF_2$ have the same frequency and phase. $RF_1$ is injected into Channel 1 of the Oscilloscope (Tektronix DPO73304D). $RF_2$ is applied to the device in the form of electromagnetic wave through a standard gain horn antenna. After that, the received signal is recorded by Channel 2 of the Oscilloscope. We measured the response delay of the ME antenna for this input signal of 540 MHz and a duration of 200 ns (this frequency has been chosen because it is the one at which the larger detection performance if the ME spin-torque microwave detector (STMD) is obtained). Supplementary Fig. S3b and Fig. S3c show an opening time delay of 12 ns and closing time delay of about 56 ns. The overall time response of the ME STMD should be comparable to the one of the ME antenna considering that the transient for the magnetization dynamics is related to the ferromagnetic resonance $f_0$ and typically is 4-5 times $1/f_0$[1].

For a comparison, we also measured the time response of conventional rod antennas as shown in supplementary Fig. S3d. Supplementary Fig. S3e and S3f display the time domain data used to estimate the opening and closing time of the rod antenna which are 12.36 ns and 15.73 ns respectively. Our experimental data show that the order of magnitude of the response time of the ME antenna is the same but with a longer closing time due to the low dissipation damping for the bulk acoustic waves at the working frequency of the ME antenna.

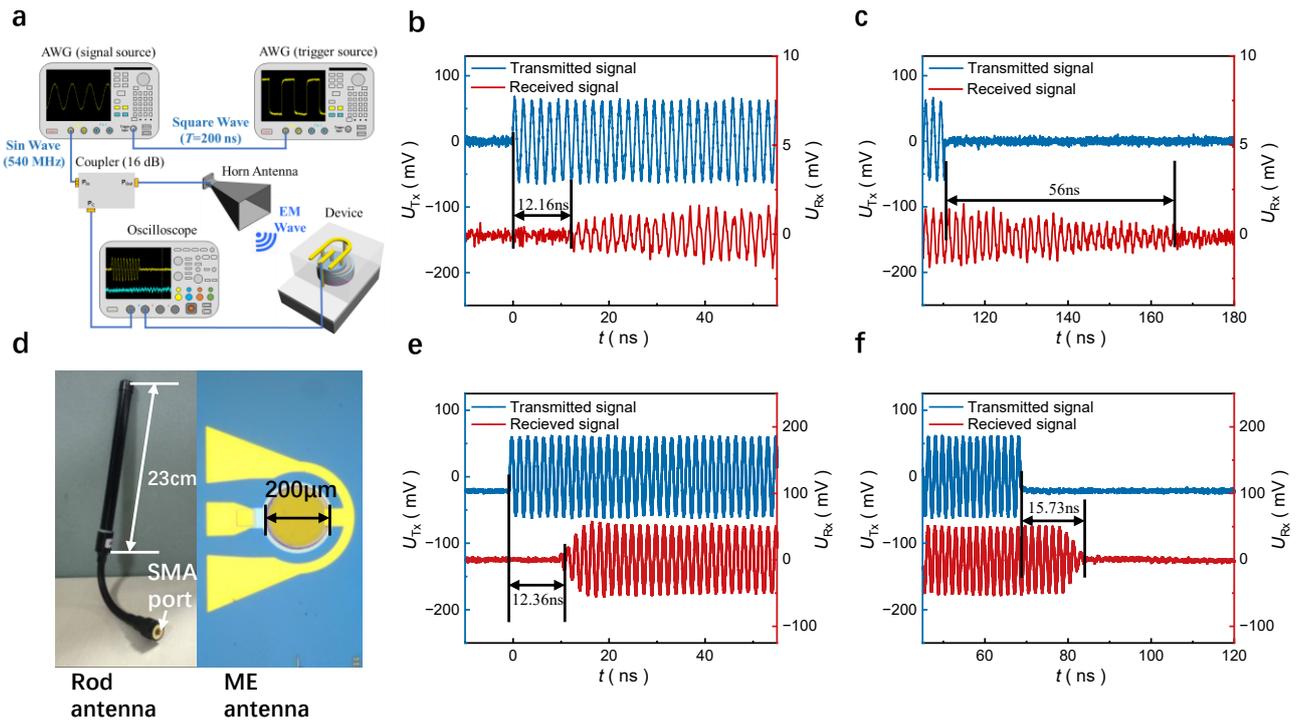

**Supplementary Figure S3 | Time response of the ME antenna. a,** Schematic diagram of the measurement system used to characterize the opening and closing delay of the ME antenna. Two AWGs are used to get a shorter period pulse sine wave. **b, c,** The open/close delay test of the ME antenna. The blue line is the original transmitted signal directly input to the oscilloscope via the coupler, and the red line is the received signal detected by the ME antenna. **d,** A comparison of a conventional rod antenna and our ME antenna. **e, f,** The open/close delay test of the rod antenna. The blue line is the original transmitted signal directly input to the oscilloscope via the coupler, and the red line is the received signal detected by the rod antenna.

**Supplementary Note 2 – Radiation pattern of the ME antenna.**

We characterized the radiation patterns of the ME antenna in a standard microwave anechoic chamber. The experimental setup is illustrated in Figure S4a. The device under test (DUT), positioned at the chamber center, was contacted via a GSG probe. A rotation stage was employed to measure the angular dependence of the radiation pattern. As shown in Figure S4b, the obtained radiation pattern exhibits well-defined main lobes and low back lobes, confirming the antenna's unidirectional characteristics. Furthermore, figure S4c and S4d reveals that the ME antenna combines the radiation characteristics of both magnetic and electric dipoles, indicating that the AlN layer and the FeGaB layer jointly contribute to the radiation of the signal[2]. For the rectification tests presented in the main text, the horn antenna was placed 1.5 meters away from the DUT, aligned along the $\theta = 270°$ and $\varphi = 90°$ indicated in Figure S4c. Based on the above radiation pattern measurements, the radiation efficiency of our antenna is 0.43%, which is similar with our previous work[3,4].

We rigorously considered the inherent parasitic radiation of the GSG probe, which could potentially interfere with measurements. To quantitatively assess this effect, the probe was modeled as a monopole antenna with a resonant frequency $f_{resonance} \approx c/4L$. Given the exposed probe length ($L$) of approximately 4 mm used in our tests, the calculated resonant frequency is about 18.75 GHz, far from the operational resonance of the ME antenna. Therefore, the probe's direct reception efficiency at the frequency of interest is exceedingly low, and its parasitic influence is considered negligible.

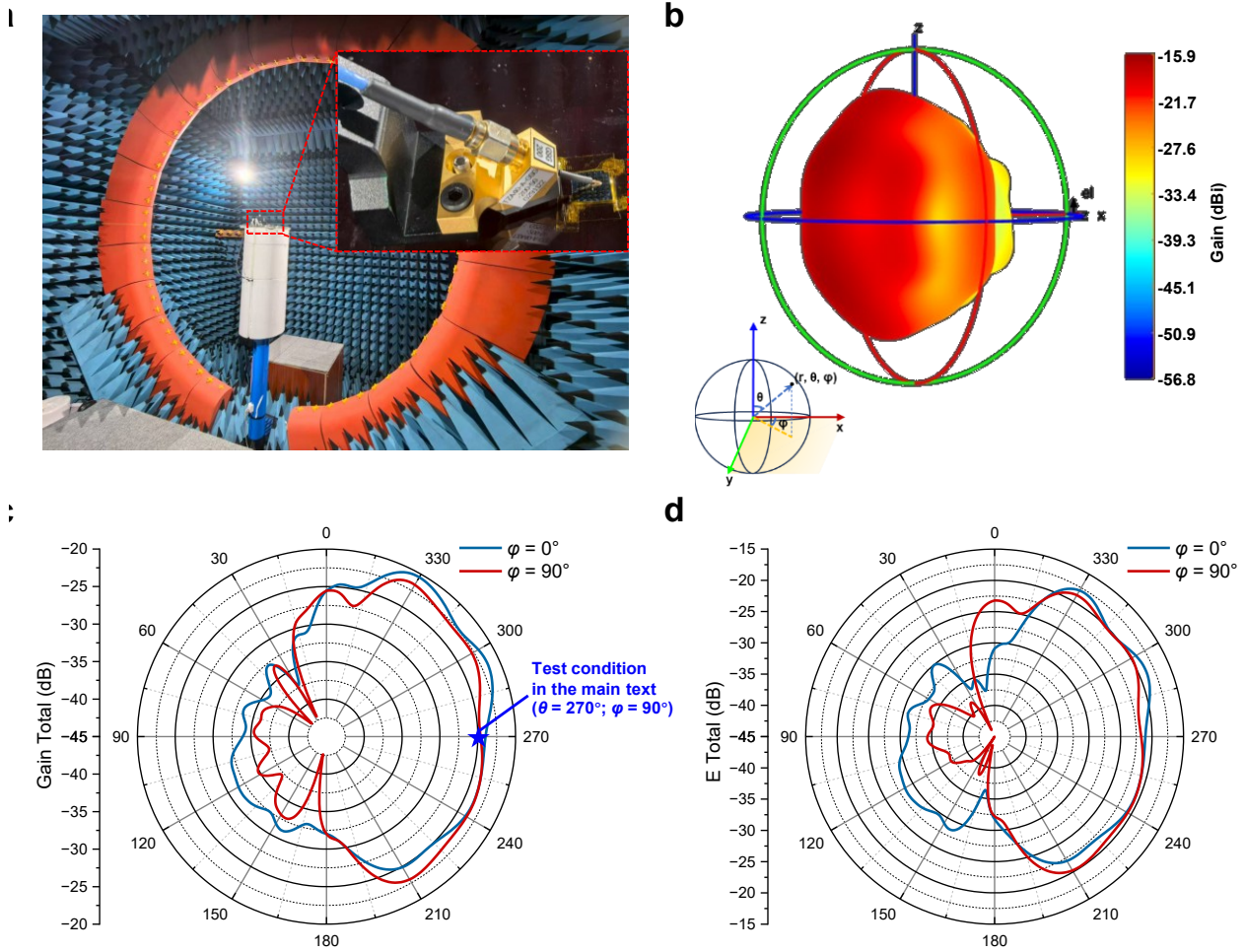

**Supplementary Figure S4 | Radiation properties of the ME antenna. a,** Schematic of the measurement setup. **b,** Three-dimensional (3D) radiation pattern. The inset at the bottom right presents the spherical coordinate system for characterization. **c,** Gain patterns in the E-plane and H-plane. In the polar coordinate plots, the angular coordinate corresponds to the θ value. **d,** Electric field patterns in the E-plane and H-plane.

**Supplementary Note 3 – Delivered power measurement.**

Supplementary Fig. S5a displays the system used to measure the amount of power generated by the horn antenna ($P_{\text{horn-antenna}}$) that is delivered to the device ($P_{\text{ME-antenna}}$). The results, obtained for a $P_{\text{horn-antenna}}$ at a frequency of 540 MHz, are summarized in Supplementary Figs. S5b and S5c where some examples of the microwave spectra measured at the ME antenna for different power emitted by the horn antenna and the relationship between the $P_{\text{horn-antenna}}$ and the $P_{\text{ME-antenna}}$ are shown respectively. Same quantitative $P_{\text{ME-antenna}}$ estimations are achieved while using a microwave power meter to evaluate the power density $D$ of the incident electromagnetic wave near the ME antenna and then multiply $D$ by its area $A$.

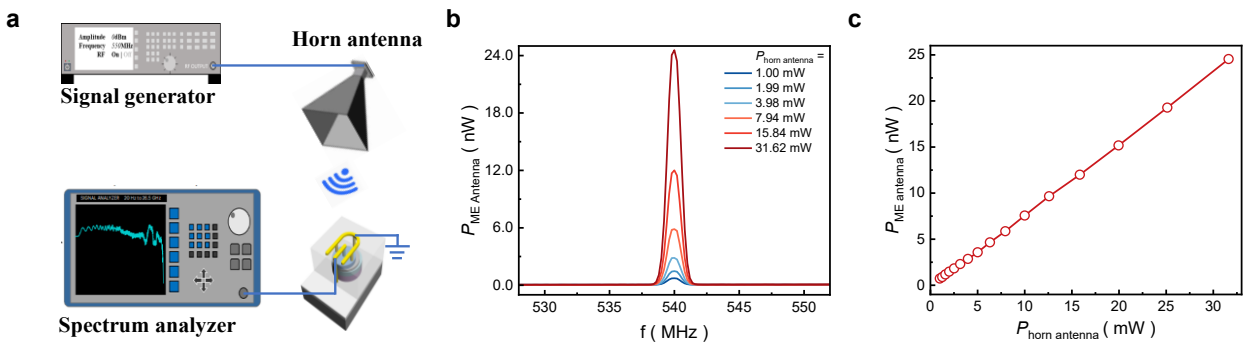

**Supplementary Figure S5 | Calibration of the input power. a,** Measurement circuit of the power calibration experiment. Due to the small area of the device, most of the emitted electromagnetic wave power $P_{\text{horn-antenna}}$ is not delivered to the device, so it is necessary to calculate the actual power received by the device $P_{\text{ME-antenna}}$. A signal of a specific frequency and power supplies the horn antenna which emits the microwave signal at the power $P_{\text{horn-antenna}}$. The ME antenna is connected to the spectrum analyzer through a probe to measure the received power $P_{\text{ME-antenna}}$. **b,** $P_{\text{ME-antenna}}$ collected by the spectrum analyzer for a 540 MHz input signal for different $P_{\text{horn-antenna}}$. **c,** Input-output relationship between $P_{\text{horn-antenna}}$ and $P_{\text{ME-antenna}}$.

**Supplementary Note 4 – Origin of the spaced rectification peaks**

As shown in Fig. 2c of the main text, the rectification response differs from the conventional MTJ ferromagnetic resonance (FMR), as the rectified signal have several equally spaced peaks. This is a signature of the high-overtone bulk acoustic resonator (HBAR) antenna[5]. The operating frequency $f$ is given by:

$$f \approx \frac{n_r v_s}{2(t_s+t_p+t_m+t_e)} \approx \frac{n_r v_s}{2t_s},$$

where $n_r$ is the harmonic order number, $v_s$ is the sound velocity in substrate, $t_s$, $t_p$, $t_m$, $t_e$ are the thicknesses of the substrate, piezoelectric layer, magnetostrictive layer and the bottom electrode respectively. Considering that the substrate thickness $t_s$ is much greater than the film thickness[4, 6]. Thus, the frequency interval between harmonics is:

$$\Delta f \approx \frac{v_s}{2(t_s+t_p+t_m+t_e)} \approx \frac{v_s}{2t_s}$$

For our device, with $t_s \approx 190$ μm and $v_s = 8433$ m/s [7], yielding theoretical $\Delta f = 22.5$ MHz, in good agreement with the measured average peak spacing of 24.7 MHz. This confirms that the rectification output results from the combined action of the MTJ and the ME antenna.

**Supplementary Note 5 – Wireless rectification detection**

The zero field results are summarized in Supplementary Fig. S6. From Fig. S6b, we can observe similar rectification peaks as in Fig. 2b, but the rectification voltage is observed at smaller dc currents 0.15 mA ~ 0.55 mA as compared to the data achieved at a bias magnetic field of 0.7 mT applied along $\theta = 15°$, with a maximum rectified voltage at 0.35 mA. The sensitivity as a function of the $P_{\text{ME-antenna}}$ is displayed in Fig. S6c. Although it is smaller than the one at 0.7 mT, it is more than 15 kV/W and still competitive with Schottky diodes.

We also tested the rectification response under negative DC bias for the device used in the Figure 2 of the main text. As shown in Supplementary Fig. S6d, the rectified signal is weaker for negative bias currents than for positive ones, consistent with the influence of DC spin-transfer-torque[8].

The rectification voltage response for different applied fields in the range -1 mT to 3 mT is summarized in Supplementary Fig. S7a. It can be observed that in the field region where large detection voltage is observed the working frequencies of the ME STMD do not exhibit significant change in this magnetic field region. Supplementary Fig. S7b shows the sensitivity as a function of the field for a $I_{\text{DC}} = 0.65$ mA and $P_{\text{ME-antenna}} = 15.1$ nW. It can be observed that the device exhibits the largest sensitivity at 0.7 mT.

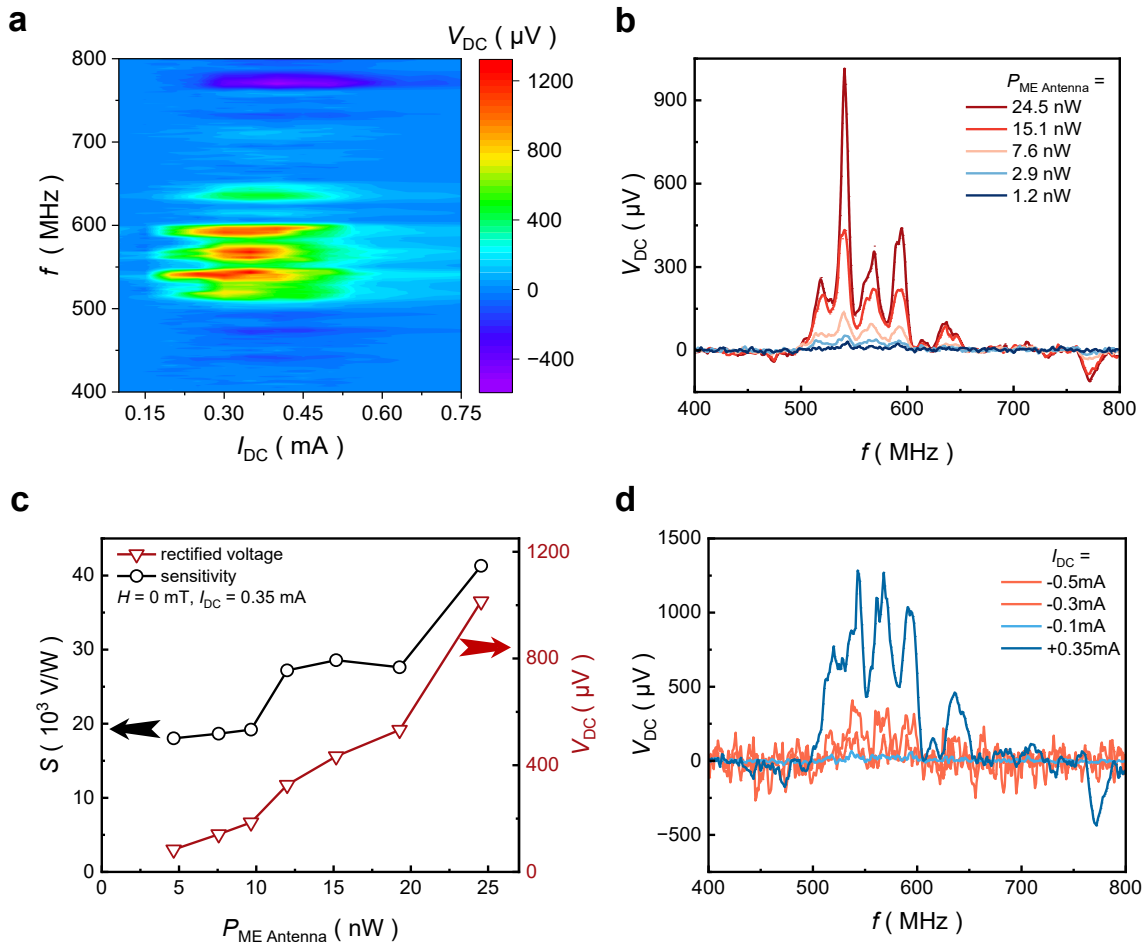

**Supplementary Figure S6 | Zero-field wireless detection. a,** Phase diagram of the rectification voltage vs. DC current (no magnetic field). Maximum rectification occurs at $I_{DC}$ = 0.35 mA. **b,** Rectification vs. frequency at different incident power levels. **c,** Detection sensitivity and rectified voltage vs. $P_{\text{ME-antenna}}$ at $I_{DC}$ = 0.35 mA. **d,** Rectification response for negative DC currents (-0.1, -0.3, and -0.5 mA) at zero magnetic field. The curve for a bias current of +0.35 mA is shown for comparison.

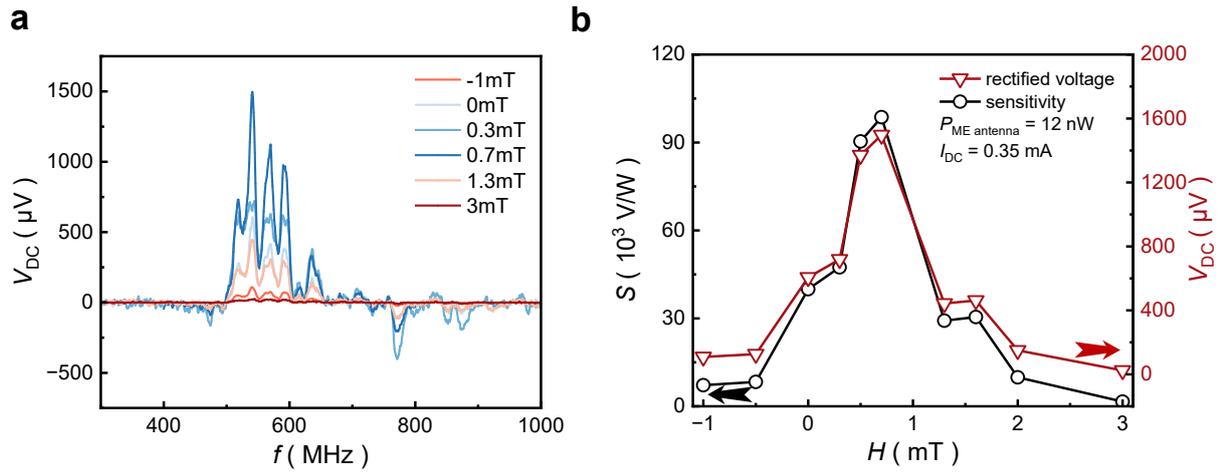

**Supplementary Figure S7 | Wireless rectification performance at different applied fields. a,** Rectified voltage response at different bias magnetic field for $I_{DC}$ = 0.65 mA and $P_{ME\text{-}antenna}$ = 15.1 nW. There is no change in the frequency of the main resonant peaks within this field region. **b,** Detection sensitivity and rectified voltage as a function of the bias magnetic field for $I_{DC}$ = 0.65 mA and $P_{ME\text{-}antenna}$ = 15.1 nW. The device exhibits a max sensitivity at $H$ = 0.7 mT.

**Supplementary Note 6 – Noise properties of the STMD**

First, we measured the nonlinear magnetic noise of the device without applying any RF signal, as shown in Fig. S8a. The device noise gradually increased with the bias current ($I_{DC}$). When an RF signal in the form of current was injected, we characterized the noise equivalent power (NEP) of the device under different power levels. The NEP increased progressively with higher injection power. However, in the actual operation of the device, it receives signals in the form of EM waves.

According to the analysis in Fig. 3 of the main text, the incident EM wave induces RF stress in the magnetoelectric heterostructure beneath the MTJ, which may potentially influence the NEP value. Therefore, we tested the noise characteristics of the device under EM wave injection ($P_{ME\ antenna}$ = 15.1 nW). The results are shown in Fig. S8d and e. Under the influence of the EM wave, the noise amplitude slightly increased. Fig. S8f displays the noise characteristics of the device under different incident EM wave power levels. This indicates that the presence of RF stress has a slight enhancement in low-frequency noise. Notably, as shown in Fig. S8b and S8d, when the DC current ($I_{DC}$) reaches 0.9 mA, the noise voltage exhibits a decrease. This behavior is likely associated with changes in magnetic moment dynamics under high current injection.

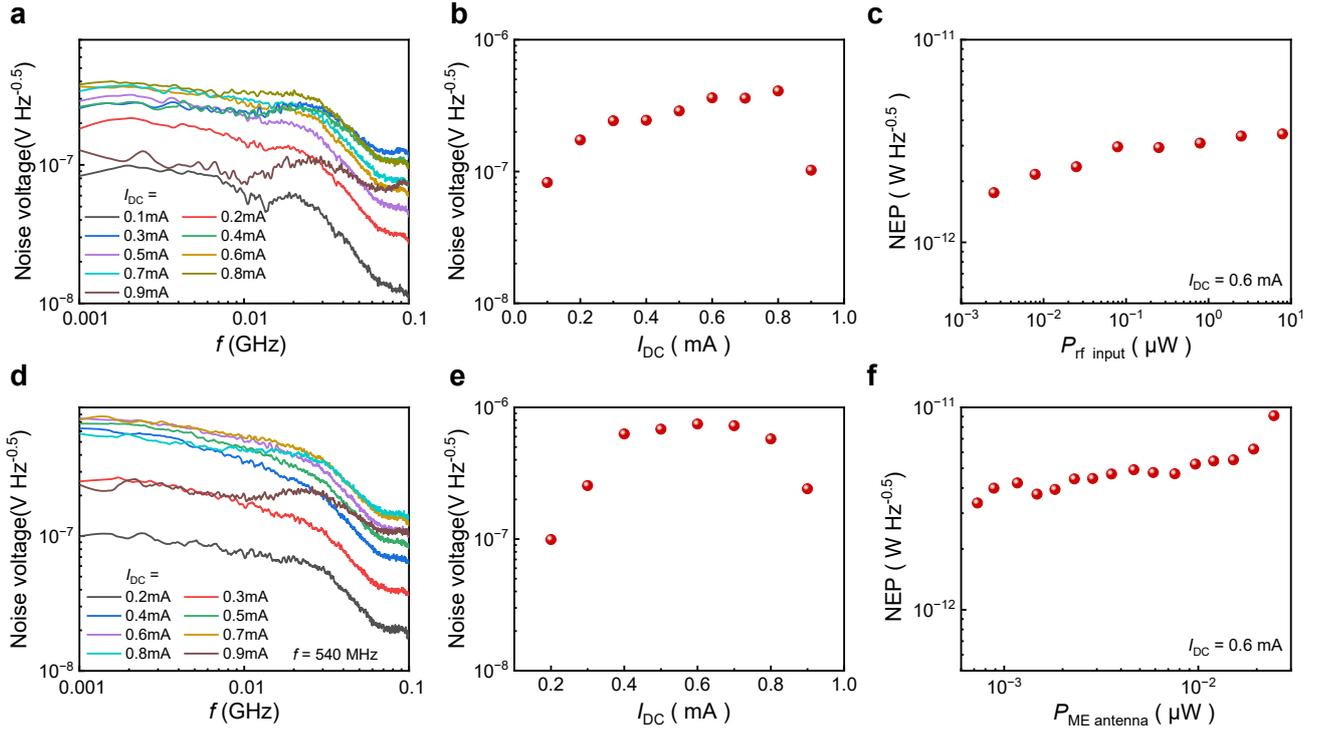

**Supplementary Figure S8 | Noise-equivalent-power (NEP). a,** Noise voltage vs. frequency for various $I_{DC}$ (no RF). **b,** Noise voltage vs. $I_{DC}$ near 0 Hz. **c,** NEP vs. RF input power ($I_{DC}$ = 0.65 mA, wired). **d,** Noise voltage vs. frequency for various $I_{DC}$ under EM wave (540 MHz). **e,** Noise voltage vs. $I_{DC}$ near 0 Hz under EM wave. **f,** NEP vs. wireless input power.

**Supplementary Note 7 – Micromagnetic simulations.**

PETASPIN, an in-house CUDA-native full micromagnetic solver, was used to perform micromagnetic simulations. This tool numerically integrates the Landau-Lifshitz-Gilbert (LLG) equation by applying the time solver scheme Adams-Bashforth [9, 10]:

$$\frac{d\boldsymbol{m}}{d\tau} = -(\boldsymbol{m} \times \boldsymbol{h}_{eff}) + \alpha_G \left(\boldsymbol{m} \times \frac{d\boldsymbol{m}}{d\tau}\right) \tag{S1}$$

where **m** = **M** / $M_S$ is the normalized magnetization of the MTJ free layer, $\alpha_G$ is the Gilbert damping, and $\tau = \gamma_0 M_S t$ is the dimensionless time, which uses $\gamma_0$ the gyromagnetic ratio and $M_S$ the saturation magnetization. The normalized effective magnetic field, **h**$_{eff}$, includes the exchange, interfacial DMI, magnetostatic, anisotropy and external fields.

The experimental MTJ is modelled as an elliptical stack as the experimental sample, with a longer in-plane dimension of $a$=420 nm and shorter in plane-dimension of $b$=260 nm. The thickness of the CoFeB free layer is 1.2 nm. The external field **H** is applied with an angle $\theta$ with the respect to the x-axis. We use a cuboidal discretization cell of 4×4×1.2 nm$^3$.

For the excitation of the STD due to the dynamics of the free layer magnetization, we consider the spin-transfer-torque (STT) from both an injected DC current density $J_{dc}$ and AC current density $J_{ac} = J_{ac,0} \sin(2\pi f_{AC} t + \varphi)$ of frequency f, phase $\varphi$ and amplitude $J_{ac,0}$. The STT term is added to Eq. (1) as:

$$\boldsymbol{\tau}_{STT} = \frac{gP\mu_B}{\gamma_0 e M_{S-FL}^2} (J_{dc} + J_{ac})[\boldsymbol{m} \times (\boldsymbol{m} \times \boldsymbol{m}_{RL})] \tag{S2}$$

where $\boldsymbol{m}_{RL}$ is the reference layer magnetization, which is set along the positive y-axis. g is the Landè factor, P is the spin-polarization equal to 0.66, $\mu_B$ is the Bohr magneton, $e$ the electron charge, and $V_{FL}$ is the volume of the free layer.

We consider exchange constant $A$ = 20 pJ/m, $M_s$ = 1200 kA/m, perpendicular anisotropy constant $K_u$ = 0.83 MJ/m$^3$, $\alpha_G$ = 0.02, and the field angle $\theta$ = 15°.

Figure S9a shows the time evolution of the normalized TMR, computed as $(1-<m_y>)/2$, where the $<m_y>$ indicates the spatially-averaged y-component of the MTJ free layer, for $J_{dc}$ = 0.5 MA/cm$^2$ and $H$ = 0.7 mT. The corresponding Fourier spectrum, illustrated in Fig. S9b, shows the existence of a peak with a broad linewidth around 400 MHz, which agrees well with the experimental data of Fig. 3a of the main text, thus confirming the excitation of incoherent magnetization dynamics. Smaller devices can exhibit microwave emissions with single peak and narrow linewidth when a uniform mode or vortex dynamics is excited [11, 12]. Figure S9c shows the critical current density to achieve the self-oscillation regime as a function of the applied field which in this field region increases, as also observed in the experiments.

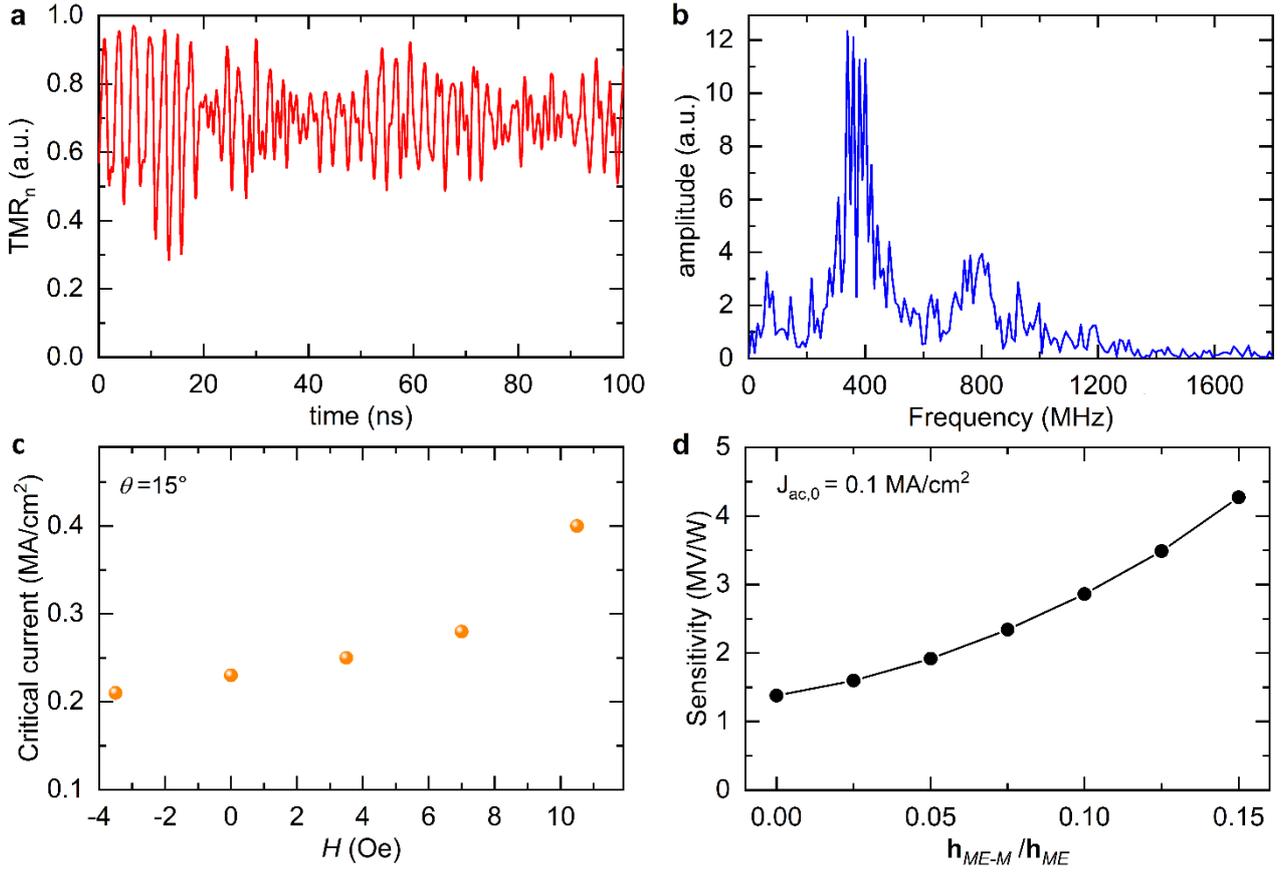

**Supplementary Figure S9 | Micromagnetic simulations results. a,** Time evolution of the normalized TMR for $J_{dc} = 0.5$ MA/cm$^2$ and $H = 0.7$ mT. **b,** Fourier spectrum of the spatially-averaged y-component of the MTJ free layer magnetization of the same simulation of panel a. **c,** Critical current density to excite the self-oscillation as a function of the applied field $H$. **d,** Sensitivity as a function of the ratio between the amplitude of the ac and dc strain ($\mathbf{h}_{ME-M}/\mathbf{h}_{ME}$) for two values of ac current (the other micromagnetic parameters are the same as panel a).

*Modeling of the strain induced by the ME antenna in the MTJ.*
This strain can be modeled as an anisotropy term to the effective field of the Landau-Lifshitz-Gilbert (LLG) in Eq. S1[13,14]:

$$\mathbf{h}_{ME} = \frac{1}{\mu_0 M_S^2}\left(\sum_{i=x,y,z} 2B_1 \varepsilon_{ii}(\mathbf{e}_i \cdot \mathbf{m})\mathbf{e}_i + \sum_{i,j=x,y,z} B_2 \varepsilon_{ij}[(\mathbf{e}_j \cdot \mathbf{m})\mathbf{e}_i + (\mathbf{e}_i \cdot)\mathbf{e}_j]\right) \quad (S3)$$

where the magnetostriction coefficients are $B_1 = -\frac{3}{2}\lambda_{100}(c_{11} - c_{12})$, and $B_2 = -3\lambda_{111}c_{44}$, with $\lambda_{100}$ and $\lambda_{111}$ being the magnetostriction constants along the [100] and [111] crystallographic directions, respectively, and $c_{11}$, $c_{12}$, and $c_{44}$ being the elastic stiffnesses. The unit of B$_1$ and B$_2$ is Jm$^{-3}$. $\mu_0$ is the vacuum permeability. $\mathbf{e}_i$ is the unit vector in the directions $i$=x,y,z. The ac strain is given by $\mathbf{h}_{ME-ac} =$

$\mathbf{h}_{ME-M}\sin(2\pi f_{AC}t + \varphi)$. For micromagnetic simulations we consider $\lambda_{100} = \lambda_{111} = \lambda = -2.5\times 10^{-5}$, $c_{11} = 0.28$ TJ/m³, $c_{12} = 0.14$ TJ/m³ and $c_{44} = 0.07$ TJ/m³ [8,9,10].

Figure S9d shows the sensitivity as a function of the ratio between the amplitude of the ac and dc strain ($\mathbf{h}_{ME-M}/\mathbf{h}_{ME}$) for $J_{ac,0} = 0.1$ MA/cm² ($J_{dc} = 0.5$ MA/cm²). These micromagnetic simulations, performed with no thermal fluctuations, confirm qualitatively the experimental results underling that the addition of an ac strain enhances the sensitivity of the ME STMD.